\documentclass[pra,twocolumn]{revtex4}
\usepackage{graphicx,bm,times}
\usepackage{tabularx} % in the preamble
\usepackage{lipsum}% http://ctan.org/pkg/lipsum-
\usepackage{amsmath}
\usepackage{gensymb}
\usepackage{bbold}
\usepackage{color,soul, mathrsfs}
\DeclareMathOperator{\tr}{tr}

\usepackage[]{natbib}
\newcommand{\e}{\text{e}}
\newcommand{\ti}{\text{i}}

\newcommand{\Hugo}[1]{{\color{black}{#1}}}
\usepackage[normalem]{ulem}

\begin{document}

\title{The Quantum-Classical Boundary for Precision Optical Phase Estimation}

\author{Patrick M. Birchall}
\email[]{patrickbirchall@gmail.com}
\author{Jeremy L. O'Brien}
\author{Jonathan C. F. Matthews}
\author{Hugo Cable}
\email[]{Hugo.Cable@Bristol.ac.uk}
\affiliation{Quantum Engineering Technology Labs, H. H. Wills Physics Laboratory and Department of Electrical and Electronic Engineering, University of Bristol, Merchant Venturers Building, Woodland Road, Bristol BS8 1FD, UK.}

\begin{abstract}
Understanding the fundamental limits on the precision to which an optical phase can be estimated is of key interest for many investigative techniques utilized across science and technology. We study the estimation of \Hugo{a fixed optical phase shift due to a sample} which has an associated optical loss, and compare phase estimation strategies using classical and non-classical probe states. \Hugo {These comparisons are based on the attainable (quantum) Fisher information calculated per number of photons absorbed or scattered by the sample throughout the sensing process.}  We find that, for a given number of incident photons upon the unknown phase, non-classical techniques in principle provide less than a $20\%$ reduction in root-mean-square-error (RMSE) in comparison with ideal classical techniques in multi-pass optical setups. Using classical techniques in a new optical setup we analyze, \Hugo{which incorporates additional stages of interference during the sensing process,} the achievable reduction in RMSE afforded by non-classical techniques falls to only $\simeq4\%$.  \Hugo{We explain how these conclusions change when non-classical techniques are compared to classical probe states in non-ideal multi-pass optical setups, with additional photon losses due to the measurement apparatus.}
\end{abstract}
\date{\today}
\maketitle

The use of non-classical techniques to perform high-precision parameter estimation is the subject of intensive research efforts \cite{giovannetti2011}. The applications of high-precision optical interferometry are widespread, and are subject to different limiting factors.  An early quantum-enhanced strategy for gravitational wave detection was proposed in 1980 using optical interferometry \cite{caves1981}. By considering the limitations placed on the system by the total power available, radiation-pressure-induced dephasing and optical loss, it was shown that squeezed states could be used to increase sensitivity.

\begin{figure}
 \includegraphics[width=1.0\columnwidth]{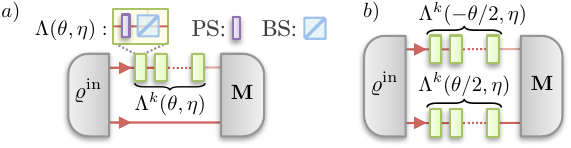}
 \caption{MP schemes using an input state $\varrho^{\text{in}}$ and $k$ passes through an absorptive or lossy phase-shift (PS), $\Lambda(\theta,\eta)$, before measurement with a positive-operator value measure (POVM).  $\Lambda(\theta,\eta)$ can be described by a beam-splitter (BS) of transmissivity $\eta$, coupling into a vacuous environment mode and a unitary phase-shift of $\theta$, $\e^{\ti\hat{n}\theta}$. $a)$ To measure a lossy phase shift in one mode, a MP strategy can be used together with a reference beam. $b)$ To measure a phase difference between two modes, equal and opposite phase shifts are considered acting on each mode \cite{jarzyna2012}.}
  \label{channel}
\end{figure}

In many subsequent treatments of interferometric quantum metrology, practical factors are set aside and the focus is on the abstract task of estimating a phase (in one mode) encoded in a state by a unitary operator $|\psi(\theta)\rangle=\exp(\ti\hat n_{1}\theta)|\psi(0)\rangle$ \cite{giovannetti2004}. The typical figure of merit is the obtainable precision, and the main result is how this scales with the number of photons used $\Delta\theta=g(N)$ \cite{holland1993}. Once dephasing and loss have been omitted, the best choice of input state for detecting small phase shifts is the N00N state $|N,0\rangle_{1,2}+|0,N\rangle_{1,2}$ \cite{dowling2008}. The use of non-classical input states allows the estimation uncertainty to be reduced from the standard quantum limit (SQL) $\Delta\theta=1/\sqrt{N}$ to $\Delta\theta=1/N$, known as the Heisenberg limit \cite{heitler1954}. Recent demonstration of violation of the SQL with the two-photon N00N state \cite{slussarenko2017} after 30 years of two-photon interference (Hong-Ou-Mandel) experiments,  has demonstrated the need to mature quantum technology (for example efficiency in photon detection and photon generation) to a sufficient level to outperform classical states in a phase estimation measurement.  However any claim to achieve a reduction in uncertainty should be understood within the context of the assumptions used to derive it. For example, if one is concerned with measuring a fixed phase instead of a time-varying one, then it is possible to improve precision beyond the SQL (as defined above) using classical states, by increasing the number of times the phase shift is applied to each photon.  This can be done using multi-pass (MP) strategies as depicted in Fig.~\ref{channel}a)  \cite{higgins2007}. Furthermore, if there is any optical loss the estimation uncertainty scaling is proportional to the SQL, and the quadratic improvement is lost \cite{knysh2011, escher2011}.

In this work we study the estimation of an optical phase shift due to a sample which remains fixed in time, where optical loss due to the sample is the dominant non-ideal process.  We focus on the precision that can be achieved whilst passing a limited number of photons through the phase-shift.  This is particularly relevant for measurements which must limit light levels to avoid altering or damaging the sample, such as measurements of biological and other delicate systems \cite{taylor2016}.  The results we report here will help development of multi-pass microscopy \cite{juffmann2016, juffmann2017}.  Consequently, we take the physical resource for our analysis to be statistical information per photon absorbed or scattered by the sample (defined below).  Our analysis assumes that optical losses (wherever they occur) always take the form of a linear loss model, which is to say that a constant fraction of the incident intensity is lost. This model applies very widely in optical experiments, though it can break down in some special circumstances (one example being two-photon absorption processes at high intensities).  We note that many applications of optical phase estimation use laser light as the probe, and rely on increases in input intensity to reduce uncertainty.  This approach remains possible until the maximum amount of incident light a sample can tolerate is reached, at which point precision gains must be achieved by increasing the amount of statistical information provided per incident photon.

For our analysis, we will compare the performance of N00N state and classical MP strategies, as well as optimal classical MP and optimal quantum strategies \cite{knysh2011, escher2011}. In search of practical quantum strategies, we then look at how well one can do with Gaussian states and intensity measurements. We will also consider classical input states in more general optical setups, to assess further scope for increased precision afforded by non-classical techniques. By investigating the potential of interferometric setups which use only classical techniques, we aim to find out what advantage can only be achieved using non-classical techniques and thereby illuminate the quantum-classical boundary for precision optical phase estimation.

We note that recent work by other authors \cite{dorner2009, demkowicz2010, demkowicz2014entangle} compares phase estimation strategies including MP strategies, on the basis of a resource which is defined as the number of particles in the probe multiplied by the number of passes through the phase shift.  Using this definition N00N states are equivalent to classical MP strategies for example.  By studying the
capabilities of {\it all} possible input states with fixed definite photon number $N$, it was found that the use of general quantum probe states instead of classical strategies can provide a reduction in RMSE of at most ~39.3\% for all values of the loss parameter involved \cite{demkowicz2014entangle}.  In contrast, the resource definition we use applies in particular for optical phase estimation, and accounts for changes in probe intensity at the sample throughout the sensing process to evaluate the total exposure of the sample.

To assess the precision capabilities of different schemes we use the Fisher information (FI) and the quantum Fisher information (QFI) \cite{helstrom1976, braunstein1994,van2000}. Once a parameter has been encoded onto a quantum state $\varrho_{\theta}=\sum_ip_i(\theta)|\psi_i(\theta)\rangle\!\langle\psi_i(\theta)|$, expressed in terms of the eigenvectors of $\varrho_\theta$, the FI is associated with a particular choice of measurement on $\varrho_\theta$.  It is defined by $F_{\mathbf{M}}(\varrho_{\theta})\!=\!\sum_i p(i|\theta)\left[\partial_\theta\!\log{p}(i|\theta)\right]^2$, associated with the probabilities of measurement outcomes $p(i|\theta) = \tr(m_i\varrho_\theta)$ for measurements on $\varrho_\theta$ with a POVM $\mathbf{M} = \{m_i\}$.  The QFI is defined as
$
\mathcal{F}(\varrho_{\theta})\equiv\tr\left[\varrho_{\theta}\mathcal{L}_{\theta}^2(\varrho_{\theta})\right],
$
where $\mathcal{L}_\theta(\varrho_{\theta})=\sum_{i,j}2\langle\psi_{i}|\frac{\partial }{\partial \theta}\varrho_\theta|\psi_{j}\rangle|\psi_{i}\rangle\!\langle\psi_{j}|\big/( p_i+ p_j)$ is the symmetric logarithmic derivative (SLD). $\mathcal{F}(\varrho_{\theta})$ and $F_{\mathbf{M}}$ bound the precision to which $\theta$ can be estimated according to the relations,
\begin{equation}
\mathcal{F}(\varrho_{\theta})\overset{2}{\geq}F_{\mathbf{M}}(\varrho_\theta) \overset{1}{\geq} 1\big/\Delta^2\theta.
\end{equation}
Inequality 1 is the regular Cram\'{e}r-Rao bound and relates  $\Delta^2\theta$, the mean-square error (MSE) of unbiased estimates of $\theta$, to the FI.  The regular Cram\'{e}r-Rao bound is saturated when many repeated estimates are performed, numbering $t$, and maximum-likelihood estimation is used i.e. $F_{\mathbf{M}}(\varrho^{\otimes t}_\theta) = t F_{\mathbf{M}}(\varrho_\theta) = 1/\Delta^2\theta$ when $t$ is large \cite{van2000}.  Inequality 2 is the Quantum Cram\'{e}r-Rao bound and it relates regular FI to the QFI. The Quantum Cram\'{e}r-Rao bound is always saturable by measuring $\varrho_\theta$ in an optimal basis i.e. $\mathcal{F}(\varrho_\theta) = \max_{\mathbf{M}} F_{\mathbf{M}}(\varrho_\theta)$ \cite{braunstein1994}. Thus QFI provides a measurement-basis independent evaluation of $\varrho_{\theta}$. By considering the input state and the setup used to encode the phase, we can make a full accounting of the resources required to achieve a given precision.

%%%%%%%%%%%%%%%%%%%%%%\section{NOON States are sub classical}
We turn now to the example of a N00N state probing a two-mode (TM) lossy phase, as illustrated in Fig.~\ref{channel}b).  \Hugo{For comparison, we consider a single-pass (SP) strategy using the N00N state versus a MP strategy using coherent states (the classical strategy).  The coherent states are assumed to have equal intensity in each mode, and initial total mean photon number of one.   An $N$-photon N00N state which has passed through the TM lossy phase once achieves QFI equal to that for the coherent states when passed through the TM lossy phase $N$ times \cite{dorner2009, demkowicz2010}.  However, the mean number of photons which pass through the phase shift are not equal in both cases. For the N00N state all $N$ photons are incident upon the phase shift. For the coherent states, the mean number of photons decreases by a factor of $\eta$ each time the probe passes through the lossy phase.  The average total number of incident photons, over all passes, is therefore $\sum_{p=1}^{N} \eta^{p-1}.$   For $\eta<1$ and $N>1$ this expression is less than $N$. Thus, the phase will absorb fewer photons using the classical strategy, and we can conclude that N00N states are actually sub-classical resources for the task we are considering.  N00N states are seen to be among the most fragile states, losing all their phase sensing capability when any number of the photons is lost. Therefore in order to find the scope for a quantum advantage, we compare classical strategies to upper bounds on the phase sensing abilities of any quantum state in the presence of loss.}
%%%%%%%%%%%%%%%%%%%%%%%%%%%%%%\section{Limits}
%%%%%%%%%%%%%%%%%%%%%%%%%%%%%%\subsection{Classical}

\emph{Classical multi-pass strategies:} Next we consider the TM case where there is no reference beam, depicted in Fig.~\ref{channel}b), which has been studied extensively in the literature.  The effect of each pass on the probe is described by application of the channel $\Lambda(\eta,-\theta/2)[\bullet]\!\otimes\!\Lambda(\eta,\theta/2)[\bullet]$, where here and throughout $\eta$ and $\theta$ are quantities intrinsic to each single application of a lossy-phase.  The actions of phase, $\theta$, and loss, $(1-\eta)$, commute \cite{demkowicz2009}.  In the ideal case when no additional loss is encountered between each application of the channel therefore, $k$ applications of the channel $\Lambda^k({\eta,\theta})[\bullet]\equiv\Lambda(\eta,\theta)[\Lambda(\eta,\theta)[\cdots\Lambda(\eta,\theta)[\bullet]\cdots]]$ perform the same operation as $\Lambda({{\eta^k},k\,\theta})$.  Hence known results for SP strategies can be easily modified for the MP case. The figure of merit we maximize is the QFI per average total number of photons lost due to all passes through the sample, which will be denoted as $\mathcal{F}^\prime$. The average total number lost at the lossy-phase $\langle\hat n_l\rangle$ over all interrogations, and the total number incident (over all interrogations) $\langle\hat n_i\rangle$, are related by
$\langle\hat{n}_l\rangle=\langle\hat{n}_i\rangle(1-\eta)$; consequently either $\langle\hat{n}_l\rangle$ or $\langle\hat{n}_i\rangle$ may be considered the resource whose use we are minimizing. Therefore $\mathcal{F}^\prime$ using $k$ passes through the phase shift is,
\begin{equation} \label{classmpplp}
\mathcal{F}^\prime(\varrho^{\text{out}},k)=\mathcal{F}(\varrho^{\text{out}}, k)\big/\big[\tr(\hat N \varrho^{\text{in}}) (1-\eta^k)\big],
\end{equation}
 where $\tr(\hat{N}\bullet)$ is the total number of photons within a state.

 For any $k$ and $\eta$, the classical (coherent) input state which maximizes $\mathcal{F}(\varrho^{\text{out}}, k)$, where $\varrho^{\text{out}}=\Lambda(\eta^k,k\theta)[\varrho^{\text{in}}]$, is the TM coherent state $\varrho^{\text{in}}_{\text{Cl}}=|\alpha/\sqrt{2},\alpha/\sqrt{2}\rangle$ with $\tr(\hat{N}\varrho^\text{in})=|\alpha|^2$. Here $\mathcal{F}(\varrho^{\text{out}}_\text{Cl},k) = \tr(\hat{N}\varrho^\text{in})\eta^kk^2$, and hence:
\begin{equation} \label{multi}
 \mathcal{F}^\prime(\varrho^{\text{out}}_{\text{Cl}},k)=\eta^kk^2\big/(1-\eta^k),
\end{equation}
which is independent of $|\alpha|^2$. MP schemes benefit from a factor of $k^2$ in the QFI as the phase shift is magnified coherently by $k$. As $\Delta^2\theta=\Delta^2 (k\tilde\theta)\implies\Delta^2\theta/k^2=\Delta^2\tilde\theta$ (which shows the MSE is reduced by a factor of $k^2$). At the same time the total transmissivity is also reduced exponentially becoming $\eta^k$.

We note that the optimal POVM for $\varrho^{\text{out}}_\text{Cl}$, such that $\mathcal{F}(\varrho^{\text{out}}_\text{Cl}) = F_{\mathbf{M}}(\varrho^{\text{out}}_\text{Cl})$, is simply a beam-splitter followed by intensity measurements. Therefore, for this strategy the achievable QFI and FI using classical techniques are equal. Due to the additional saturability of the the regular Cram\'{e}r-Rao bound, a precision of $ \mathcal{F}(\varrho^{\otimes t}) = 1/\Delta^2\theta$ is achievable when the number of estimates $t$ is large. Therefore, when the average total number of lost photons $l = t \times \tr(\hat N \varrho^\text{in})(1-\eta) $ is large an achievable precision is $\frac{1}{\Delta\theta} =\sqrt{l \mathcal{F}^\prime(\varrho)}$ since $\mathcal{F}(\varrho^{\otimes t}) = l \mathcal{F}^\prime(\varrho)$.

 In order to find an analytic solution for the optimal number of passes $k$ we permit $k$ to be non-integer. \Hugo{Varying the number of passes here can have the interpretation of varying the thickness of the sample used, where probe intensity is attenuated exponentially with increasing thickness.  $ \mathcal{F}^\prime(\varrho^{\text{out}}_{\text{Cl}}, k)$ has a single peak over positive values for $k$, and tends to zero as $k$ tends to zero or infinity.  By setting the derivative with respect to $k$ to 0, we find} $ \mathcal{F}^\prime(\varrho^{\text{out}}_{\text{Cl}}, k)$ is optimized when:
$
 k_{\text{opt}}=-(2+\mathcal{W})\big/\ln(\eta)\approx-1.59/\ln(\eta),
$
where $\mathcal{W}\equiv_{}W_0(-2/\e^2)\approx-0.406$ is the main branch of the Lambert W function at $-2/\e^2$ \cite{corless1996}. \Hugo{Using the defining equation for the Lambert function, $z=W(z)e^{W(z)}$ for all complex numbers $z$, we can write $\eta^{k_{\text{opt}}}=\e^{-(2+\mathcal{W})}=-\frac{\mathcal{W}}{2}$}.  This solution for $k$ keeps the overall proportion of light transmitted constant $\eta^{k_{\text{opt}}}\approx_{}20.32\%$. Counterintuitively, we have found that in order to lose as few photons overall, it is best for the lossy phase to lose $\sim\!80\%$ of the total input light. In which case,
\begin{equation} \label{optcl}
 \mathcal{F}^\prime(\varrho_{\text{Cl}}^{\text{out}},  k_{\text{opt}})= -\mathcal{W}(2+\mathcal{W})\big/\ln^2(\eta)\approx 0.648/\ln^2(\eta).
\end{equation}
\Hugo{$k_{\text{opt}}$ decreases towards $0$ as $\eta$ decreases, and $k_{\text{opt}}\!\!\gg\!\!1$ for highly-transmissive samples.}
 \begin{figure*}
 \includegraphics[width=0.99\textwidth]{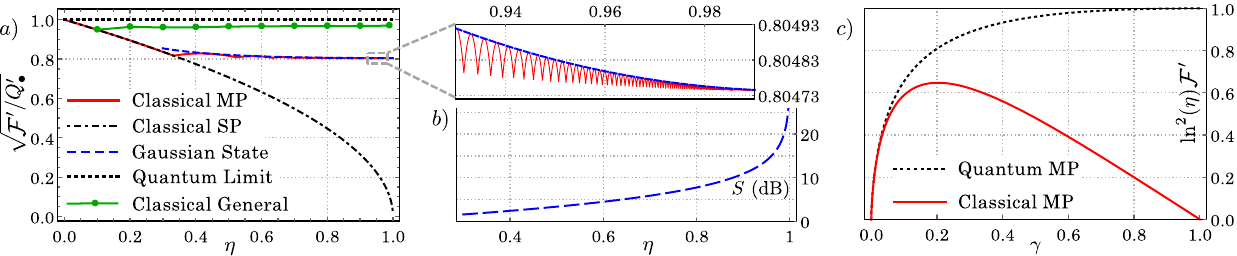}
 \caption{$a)$ Ratio for the achievable precision $1/\Delta\theta=\sqrt{l\mathcal{F}^\prime(\varrho)}$ for each scheme discussed normalized to the relevant SM or TM quantum limit on precision $1/\Delta\theta_Q=\sqrt{lQ^\prime_\bullet},\,\bullet=\text{SM}$ or $\text{TM}$.  The corresponding quantum limits assume an equal mean number of lost photons, $l$.  A discrete number of interrogations $k$ is used,  and the magnified inset shows bumps each corresponding to a different integer $k$.  $b)$ Squeezing, in dB, required for the phase-squeezed Gaussian state shown in the left most plot. $c)$ $\mathcal{F}^\prime$ for classical or optimal quantum states when interrogating a phase with a transmissivity of $\eta$ in a MP configuration. Classical-MP strategies require lower overall transmission, $\gamma=\eta^k$, to perform well. The classical MP curve remains relatively flat near the optimal $\gamma$, hence $k$ does not need to be set precisely for the classical MP scheme to perform well.}
\label{plots}
\end{figure*}

%%%%%%%%%%%%%%%%%%%%%%\subsection{Quantum}
Now we consider the best performance achievable using any quantum input state in a two-mode (TM) setup (as in Fig.~\ref{channel}b), where again each application of the channel is given by $\Lambda(\eta,-\theta/2)[\bullet]\!\otimes\!\Lambda(\eta,\theta/2)[\bullet]$.  For the SP strategy the QFI is bounded above, for any input state $\varrho^{\text{in}}$ and transmissivity, by \cite{knysh2011,escher2011, demkowicz2013}:
\begin{equation} \label{twomodebound}
 Q_{\text{TM}}\equiv\tr(\hat N\varrho^\text{in})\eta\big/(1-\eta) \geq \mathcal{F}(\varrho^{\text{out}}).
\end{equation}
This bound was derived in Ref.~\cite{knysh2011} and Ref.~\cite{escher2011} for states with a definite but arbitrary photon number, and was shown to be saturable in the large photon number limit in Ref.~\cite{knysh2011}. Subsequently, the bound in Eq.~(\ref{twomodebound}) was shown to apply to states with an indefinite and arbitrary photon number in Ref.~\cite{ demkowicz2013} when no additional reference beams are used (as we are considering in this section). The bound in Eq.~(\ref{twomodebound}) is saturable only when the photon number variance in each mode is divergent, and Ref.~\cite{escher2011} also provides a tighter bound which constrains states with finite number variance in each arm. As the bound in  Eq.~(\ref{twomodebound}) is asymptotically saturable for large $\tr(\hat N\varrho^\text{in})$, this leads to a saturable bound on the QFI per lost photon of: $Q^\prime_{\text{TM}}=\eta\big/(1-\eta)^2$. For MP strategies the bound becomes:
$Q^\prime_{\text{TM}}(k)=\eta^k k^2/(1-\eta^k)^2$ as described in Appendix A. This expression is maximized in the limit $\lim_{k\rightarrow0}Q^\prime_{\text{TM}}(k)=1\big/\ln^2(\eta)$, where $k$ is permitted to be continuous to allow for a fair comparison with the previous classical strategy.
%%%%%%%%% REFERENCE BEAM
Therefore, we find the ratio of achievable precisions for a large mean number of incident photons to be:
\begin{equation} \label{ratio}
\frac{\Delta\theta_{\text{Q}}}{\Delta \theta_{\text{Cl}}}=\sqrt{\frac{ \mathcal{F}^\prime(\varrho^{\text{out}}_{\text{Cl}},k_{\text{opt}})}{\lim_{k\rightarrow 0} Q^\prime_{\text{TM}}(k)}}\approx 0.805,
\end{equation}
i.e. the reduction in RMSE afforded by non-classical input states over classical MP strategies is $\sim\!19.5\%$.  In the limit of $k\!\rightarrow0$ the amount of information gained by the quantum strategy goes to zero. This must be compensated by increased probe intensity (so that a fixed number of photons are lost) to obtain a given amount of information.

If instead the lossy phase is confined to a single mode (SM), as in Fig.~\ref{channel}a), then a reference beam can provide additional precision. The optimal classical state now comprises two coherent states $\varrho_{\text{Cl}}=|\alpha,\beta\rangle$ where $|\beta|$ is very large. In this situation $\mathcal{F}^{\prime}$ can be calculated as above to be four times larger than the expression given in Eq.~(\ref{multi}), and is maximized for the same value of $k$. Quantum strategies can also benefit from a reference beam. In this case the bound on quantum strategies is four times larger than given in Eq.~(\ref{twomodebound}), and the mean number of photons in the input state $\tr(\hat N\varrho^{\text{in}})$ is replaced by those which are incident upon the lossy phase, $\langle\hat n_i\rangle$, giving a revised quantum bound on the QFI per lost photon, $Q_{\text{SM}}=4 \langle\hat n_i \rangle \eta\big/(1-\eta)$ \cite{escher2011}, and associated optimal MP QFI per lost photon for continuous $k$:
\begin{equation} \label{quantum_optimal}
\textstyle{\lim_{k\rightarrow0}Q^\prime_{\text{SM}}(k)=4\big/\ln^2(\eta)}.
\end{equation}
The ratio between quantum and classical strategies remains unchanged. As stated in Ref.~\cite{escher2011} the bound for SM phase estimation, $Q_{\text{SM}}$, has tremendous universality, and applies even when arbitrary operations are performed between interrogations of the phase.  Moreover it too is saturable in the large $\langle_{}n_i\rangle$ limit e.g. a highly squeezed-vacuum input state \cite{aspachs2009}.

When measuring the phase shift imparted by a gas or a liquid medium, the length of the sample, corresponding to a choice of $k$, may be chosen from a continuum of values; however if the system under investigation is intrinsically discrete then $k$ must be an integer to reflect this. Restricting $k_\text{opt}$ to be an integer number of passes alters the ratio in Eq.~(\ref{ratio}). The effect of this restriction is plotted in Fig.~\ref{plots}a) and shows that the reduction in RMSE granted by non-classical techniques remains less than $20\%$ for any value of $\eta$ when a discrete number of passes must be used.

%%%%%%%%%%%%%%%%%%%%%%%%%%%%%%%%%%%%%%%%%%%%
%%%%%%%%%%%        FIXED AMOUNT OF DECOHERENCE         %%%%%%%%%%%%%
%%%%%%%%%%%%%%%%%%%%%%%%%%%%%%%%%%%%%%%%%%%%

%%%%%%%%%%%%%%%%%%%%%%%%%%%%%%%%%%%%%%%%%\subsection{Fixed Amounts of loss}
\emph{Optimal strategies:} So far it has been shown that the optimal number of passes for both quantum and classical input states result in a fixed amount of loss. In fact, we will now argue that this is true for more general optical setups.   The setups we consider comprise $h$ multi-pass modules of $k$ lossy phases, each followed by an adaptive control phase $\phi$. The corresponding quantum channel is $\mathcal{E}_h\{\Lambda(\eta^k,k\theta+\phi)[\cdots\mathcal{E}_1\{\Lambda(\eta^k,k\theta+\phi)[\bullet]\}\cdots]\}$, where $\mathcal{E}_i\{\bullet\}$ describes an arbitrary quantum channel enacted by the optical setup (between the $i^{\text{th}}$ and $i+1^{\text{th}}$ multi-pass modules). This network will produce an output state, $\varrho(\eta^k,k\theta+\phi)=\sum_{i}p_i|\psi_i\rangle\!\langle\psi_i|$, where all the terms may depend on $\eta^k$ and $k\theta+\phi$.  The general form for QFI per lost photon is then given by,
\begin{equation} \label{fisherform1}
\begin{split}
\mathcal{F}^{\prime}\left[ \varrho(k\theta+\phi,\eta^k)\right]&=\tr\left[\varrho\mathcal{L}_{\theta}^2(\varrho)\right]\big/\tr\left[\hat N(\varrho-\varrho^{\text{in}})\right],\\
\mathcal{L}_{\theta}(\varrho)=k\textstyle{\sum_{i,j}2\langle\psi_{i}}&\textstyle{|\frac{\partial }{\partial_{}k\theta+\phi}(\varrho)|\psi_{j}\rangle|\psi_{i}\rangle\!\langle\psi_{j}|\big/(p_i+p_j)}.
\end{split}
\end{equation}
Since the SLD $\mathcal{L}_\theta(\varrho)$ is a function of $k$, $k\theta+\phi$ and $\eta^k$, the QFI per lost photon is of the form $\mathcal{F}^\prime(\varrho)=k^2 f(k\theta+\phi,\eta^k)$ for some function $f$.
By re-parameterizing with $\eta^{k}=\gamma$ and $k\theta+\phi=\varphi$ we see that $k=\ln\gamma/\ln\eta$ and,
\begin{equation} \label{fisherform2}
\mathcal{F}^{\prime}(\varrho)=\ln^{-2}\eta\,[\ln^{2}\gamma\,f(\varphi,\gamma)].
\end{equation}
 Thus optimizing $\mathcal{F}^{\prime}(\varrho)$ over $k$ and $\phi$ for all values of $\eta$ and $\theta$ corresponds to optimizing $\ln^{2}\gamma\,f(\varphi,\gamma)$ over $\varphi$ and $\gamma$. Whilst $\mathcal{F}^{\prime}$ may have very complex dependence on $\varphi$, $\gamma$ and each $\mathcal{E}_i$, the following statement always applies for any setup comprised of $h$ multi-pass modules: $k$ and $\phi$ should be adjusted to produce the same output state independently of the loss encountered with each individual pass of the state through the channel, i.e. there is an optimal amount of total loss that the state should undergo.

 We note that all setups in which there is at least a single pass though a lossy phase can be considered to contain a multi-pass module, with a $k$ of at least one. If $k$ is restricted to be an integer then it will not always be possible to achieve a desired amount of loss and the strategy will be degraded in comparison with the optimal continuous-$k$ strategy. This effect will be more significant when $\eta$ is small and there is little freedom to choose how much light is lost.

 \Hugo{For the general optical setups we consider, note that when there is no prior information about $\theta$, $\varphi$ cannot initially be controlled to a high precision. As more probe states are used, information about $\theta$ is gained, and the precision to which $\varphi$ can be set by adjusting $\phi$ can be improved. This continual adjustment ensures that the procedure obtains measurement outcomes which are highly sensitive to $\theta$. Over many repetitions of the measurement, the fraction of information gained using a significantly suboptimal value for $\varphi$ can be continually reduced. The QFI for all of the strategies we discuss grows proportional to the number of photons in the  input state; therefore the requirement for many repetitions will not reduce the performance of these strategies when the total number of photons used is large. In the limit of $\eta\rightarrow1$ our analysis breaks down since the optimal values for $k$ tends to infinity. Evaluating the merit of different strategies when $\eta\simeq1$ requires a Bayesian analysis \cite{berry2009}.  We note however that no incident power is absorbed in this limit, and therefore the absorbed power cannot damage the sample under investigation.}

Applying the parameterization $\gamma=\eta^k$ used to find Eq.~(\ref{fisherform2}), the FI per incident photon of the classical- and quantum-MP strategies can be compared as a function of $\gamma$ as shown in Fig.~\ref{plots}c). A comparison of the RMSEs associated with the classical- and quantum-MP strategies corresponds to the square root of lines in Fig.~\ref{plots}c). This plot shows that, whilst classical-MP strategies require $\gamma = \eta^{k_{\text{opt}}}\simeq 20.32\%$ to perform optimally, the classical MP curve remains relatively flat near $k_{\text{opt}}$ hence $k$ only needs to be set such that $\gamma$ is near $20.32\pm10\%$.

%%%%%%%%%%%%%%%%%%%%%%%%%%%%%%%%%%%%%%%%%\section{Practicality of achieving these bounds}
\emph{Gaussian-state strategy:}  For SP-TM phase estimation with balanced loss in each of the two modes, the problem of maximizing precision whilst passing a given mean number of photons through a sample, is equivalent to maximizing precision for a fixed mean number of photons in the input state. This is a task which as been investigated thoroughly \cite{dorner2009, demkowicz2009, kacprowicz2010, thomas2011, cable2010}. It has been shown for TM lossy phase estimation that the Gaussian state suggested in \cite{caves1981}, coupled with intensity measurements, saturates the ultimate limit on precision given in Eq.~(\ref{twomodebound}) \cite{caves1981,demkowicz2013}. This choice of probe state is implemented in GEO600, which is an interferometer for gravitational-wave detection \cite{ligo2011,grote2013,aasi2013}. However, new analysis is needed to determine optimal strategies for both SM phase estimation and TM phase estimation with unbalanced loss: taking the resource to be the number of lost photons due to the sample rather than the total number of photons in an optical state (including any required reference beam) leads to different solutions. In light of this, we now evaluate the potential for practical gains using non-classical states for SM phase estimation with SM Gaussian input states and homodyne detection.  Our theoretical comparison is especially important as these techniques are relatively simple to implement experimentally.

SM Gaussian states are fully characterized by a displacement vector $\bm{d}$ of means, $d_i=\langle\hat x_i\rangle$, and matrix $\mathbf{\Gamma}$ of covariances, $\Gamma_{ij}=\frac{1}{2}\langle\hat{x}_i\hat{x}_j+\hat{x}_j\hat{x}_i \rangle-\langle\hat{x}_i\rangle\langle\hat{x}_j\rangle$, of the quadrature operators $\hat{x}_1=\frac{1}{2}(\hat{a}^{\dagger}+\hat{a})$ and $\hat{x}_2=\frac{1}{2}i(\hat{a}^{\dagger}-\hat{a})$ \cite{loudon1987}. Homodyne measurement of the SM state using a reference beam can be performed to measure the $\hat x_1$ quadrature \cite{loudon1987}. An arbitrary SM pure Gaussian state can be defined by the squeezing $\hat S(r,\phi)=\exp[\frac{1}{2}r({\e}^{\ti \phi}\hat{a}^2-{\e}^{-\ti \phi}{\hat{a}^{\dagger2}})]$, displacement $\hat D(\alpha)=\exp[\alpha(\hat{a}^{\dagger}-\hat{a})]$, and rotation $\hat{R}(\varphi)=\exp(\ti  \hat{a}^{\dagger}\hat{a}\,\varphi)$ operators acting on a vacuum state:
$\hat{R}(\varphi)\hat D(\alpha)\hat S(r,\phi)|0\rangle$.  All arguments here are real and the mean number of photons is
$\langle\hat{N}\rangle=\alpha^2+\sinh^2(r)$.
For phase-squeezed coherent states, $\phi=\pi$, we evaluate the FI of the output state $\varrho^{\text{out}}_{\text{G}}$ at the homodyne angle $\varphi=\pi/2-\theta$ to be,
\begin{equation}
F_{\hat x}(\varrho^{\text{out}}_{\text{G}})\big|_{\varphi=\pi/2-\theta}=4\alpha^2\eta\big/\left[1+(e^{-2r}-1)\eta\right],
\end{equation}
where a full derivation is provided in appendix B.  When the number of photons contributing to the initial squeezing $n_{\text{sq}}=\sinh^2(r)$ is fixed but the total number of input photons is large, this scheme achieves a FI per lost photon of:
\begin{equation}\label{gauss}
 \lim_{\langle\hat{N}\rangle\rightarrow\infty}F_{\hat{x}}^\prime(\varrho^{\text{out}}_{\text{G}})=\frac{1-\eta }{2\eta_{}n_{\text{sq}}-2\eta\sqrt{n_{\text{sq}}(n_{\text{sq}}+1)}+1}\frac{4\eta}{(1-\eta)^2}.
\end{equation}
Recalling the corresponding quantum bound $Q^{\prime}_{\text{SM}}=4\eta/(1-\eta)^2$, we see that for a fixed $n_{\text{sq}}$ and a given value of $\eta$, a constant fraction of $Q^{\prime}_{\text{SM}}$ can be achieved for a large total number of photons. Moreover Eq.~(\ref{gauss}) saturates $Q^{\prime}_{\text{SM}}$ for large $n_{\text{sq}}$.
We can also use Eq.~(\ref{gauss}) to find the amount of squeezing in dB, $S=20 r/\ln(10)$, needed to gain as much FI per lost photon as the optimal SM classical MP strategy (which is given by four times the expression in Eq.~(\ref{optcl})). This is plotted in Fig.~\ref{plots}b).

%%%%%%%%%%%%%%%%%%%%%%%%%\section{More general networks}
 \begin{figure}
 \includegraphics[width=0.82\columnwidth]{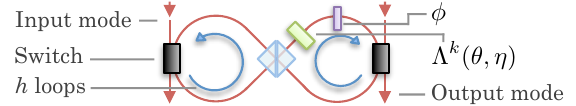}
 \caption{A setup which  interrogates an absorbing phase shift $h$ times, with interference between each interrogation.}
 \label{general_setups}
\end{figure}

\emph{Improved classical strategies:} Next we ask if classical-MP strategies, which contain a single multi-pass module in the terminology of {\it optimal strategies} above, can be surpassed by more complicated classical strategies. We consider the interrogation of a SM lossy phase embedded in a setup for which, between each interrogation of the sample, interference occurs between the outgoing mode from the phase and another optical mode.  This is illustrated in Fig.~\ref{general_setups}. Full details for this setup are provided in appendix C.  We find using a heuristic optimization for a series of values of $\eta = 0.1,0.2...0.9$ that the optimal QFI per lost photon exceeds the classical MP strategy, and the scope for quantum enhancement is further reduced to an average reduction in RMSE of $3.6\%$ (over this range of $\eta$). Individual results are plotted in Fig.~\ref{plots}a), and details of the optimization can be found in appendix D. The optimizations are not exhaustive and there may be further room for improvement. Using additional reference beams, the same improvement can be made for TM phase estimation (as discussed in appendix E).

%%%%%%%%%%%%%%%%%%%%%%%%%\section{More loss}
{\it Discussion:} By carefully considering resources, we have investigated the differences between ideal classical and ideal quantum strategies to illuminate any fundamental advantage which can only be obtained by using non-classical techniques. Our analysis is also instructive for practical situations when the dominant non-ideal effect is optical loss encountered at the lossy-phase; as technology improves over time and losses in experimental components are reduced, this regime will extend to lossy phases with ever higher transmissivities. In appendix F we analyze a non-ideal multi-pass setup which has additional reduced transmissivities $\eta_{{p}}$, $\eta_{r}$ and $\eta_{m}$ associated with state-preparation, each single round-trip between applications of the lossy phase, and the measurement stage respectively. We establish that if the round-trip loss is less than the combined state preparation and measurement loss ($\eta_{r}\geq \eta_{p}\eta_{m}$), the scope for reduced RMSE from non-classical techniques remains below $20\%$. The regime ($\eta_{r}\geq \eta_{p}\eta_{m}$) is already practically relevant, applying to GEO600 where $\eta_{p}\eta_{m} \simeq 0.44$ \cite{aasi2013}, since low loss $(\leq10^{-4})$ \footnote{For example, commercially available Semrock High-reflectivity Mirrors.} passive optical elements can lead to low round-trip loss (whilst state-preparation and measurement require more complex and less-efficient components such as non-linear crystals, filters and photo-detectors). Conversely when $\eta_{p}\eta_{m}\eta\simeq1$ and $\eta_{r} \ll 1$, quantum strategies operate near ideally whilst classical strategies cannot perform comparably (as use of large $k$ is effectively prohibited by high round-trip loss). The performance of the improved classical strategy will be more heavily degraded by imperfect components due to its increased complexity and this may nullify any precisions gains that might be achievable in principle.

%%%%%%%%%%%%%%%%%%%%%\section{Conclusions}
In conclusion, we have found that for a given number of lost or absorbed photons used to measure a fixed optical phase, the precision increase that can be achieved by using non-classical techniques is small. By explicitly considering resources, we have also argued that SM phase estimation requires different estimation strategies, and provided a practical quantum strategy based on Gaussian states. This practical quantum strategy could be used for the estimation of a phase which is rapidly varying in time, in contrast to classical techniques which suffer from blurring of temporal features due to increased interrogation time. An example of TM phase estimation constrained by temporal variations in the phase shift is gravitational-wave interferometry. Gravitational-wave detectors operate in a regime where light is stored by cavities within the arms of an interferometer for the maximum time possible whilst achieving their desired temporal resolution \cite{pitkin2011}. In cases where the interrogation time cannot be further increased, there can be a big difference between the precision achievable using classical versus non-classical techniques.

\begin{acknowledgments}
\textbf{Acknowledgments}: We would like to thank D. Mahler for helpful comments on this manuscript. This work was supported by EPSRC, ERC, PICQUE, BBOI, US Army Research Office (ARO) Grant No. W911NF-14-1-0133, U.S. Air Force Office of Scientific Research (AFOSR) and the Centre for Nanoscience and Quantum Information (NSQI). J.L.O'B. acknowledges a Royal Society Wolfson Merit Award and a Royal Academy of Engineering Chair in Emerging Technologies. J.C.F.M. and J.L.O'B acknowledge fellowship support from the Engineering and Physical Sciences Research Council (EPSRC, UK).  Data access statement: supporting data are provided within this paper.  Plots are generated from analytic functions and numerical searches described in the paper and appendices.
\end{acknowledgments}

\bibliography{classicalbib}

\begin{thebibliography}{35}
\expandafter\ifx\csname natexlab\endcsname\relax\def\natexlab#1{#1}\fi
\expandafter\ifx\csname bibnamefont\endcsname\relax
  \def\bibnamefont#1{#1}\fi
\expandafter\ifx\csname bibfnamefont\endcsname\relax
  \def\bibfnamefont#1{#1}\fi
\expandafter\ifx\csname citenamefont\endcsname\relax
  \def\citenamefont#1{#1}\fi
\expandafter\ifx\csname url\endcsname\relax
  \def\url#1{\texttt{#1}}\fi
\expandafter\ifx\csname urlprefix\endcsname\relax\def\urlprefix{URL }\fi
\providecommand{\bibinfo}[2]{#2}
\providecommand{\eprint}[2][]{\url{#2}}

\bibitem[{\citenamefont{Giovannetti et~al.}(2011)\citenamefont{Giovannetti,
  Lloyd, and Maccone}}]{giovannetti2011}
\bibinfo{author}{\bibfnamefont{V.}~\bibnamefont{Giovannetti}},
  \bibinfo{author}{\bibfnamefont{S.}~\bibnamefont{Lloyd}}, \bibnamefont{and}
  \bibinfo{author}{\bibfnamefont{L.}~\bibnamefont{Maccone}},
  \bibinfo{journal}{Nature Photon.} \textbf{\bibinfo{volume}{5}},
  \bibinfo{pages}{222} (\bibinfo{year}{2011}).

\bibitem[{\citenamefont{Caves}(1981)}]{caves1981}
\bibinfo{author}{\bibfnamefont{C.~M.} \bibnamefont{Caves}},
  \bibinfo{journal}{Phys. Rev. D.} \textbf{\bibinfo{volume}{23}},
  \bibinfo{pages}{1693} (\bibinfo{year}{1981}).

\bibitem[{\citenamefont{Giovannetti et~al.}(2004)\citenamefont{Giovannetti,
  Lloyd, and Maccone}}]{giovannetti2004}
\bibinfo{author}{\bibfnamefont{V.}~\bibnamefont{Giovannetti}},
  \bibinfo{author}{\bibfnamefont{S.}~\bibnamefont{Lloyd}}, \bibnamefont{and}
  \bibinfo{author}{\bibfnamefont{L.}~\bibnamefont{Maccone}},
  \bibinfo{journal}{Science} \textbf{\bibinfo{volume}{306}},
  \bibinfo{pages}{1330} (\bibinfo{year}{2004}).

\bibitem[{\citenamefont{Holland and Burnett}(1993)}]{holland1993}
\bibinfo{author}{\bibfnamefont{M.~J.} \bibnamefont{Holland}} \bibnamefont{and}
  \bibinfo{author}{\bibfnamefont{K.}~\bibnamefont{Burnett}},
  \bibinfo{journal}{Phys. Rev. Lett.} \textbf{\bibinfo{volume}{71}},
  \bibinfo{pages}{1355} (\bibinfo{year}{1993}).

\bibitem[{\citenamefont{Dowling}(2008)}]{dowling2008}
\bibinfo{author}{\bibfnamefont{J.~P.} \bibnamefont{Dowling}},
  \bibinfo{journal}{Contemp. Phys.} \textbf{\bibinfo{volume}{49}},
  \bibinfo{pages}{125} (\bibinfo{year}{2008}).

\bibitem[{\citenamefont{Heitler}(1954)}]{heitler1954}
\bibinfo{author}{\bibfnamefont{W.}~\bibnamefont{Heitler}},
  \emph{\bibinfo{title}{The Quantum Theory of Radiation, 3rd ed.}}
  (\bibinfo{publisher}{Oxford University Press, London}, \bibinfo{year}{1954}).

\bibitem[{\citenamefont{Slussarenko et~al.}(2017)\citenamefont{Slussarenko,
  Weston, Chrzanowski, Shalm, Verma, Nam, and Pryde}}]{slussarenko2017}
\bibinfo{author}{\bibfnamefont{S.}~\bibnamefont{Slussarenko}},
  \bibinfo{author}{\bibfnamefont{M.~M.} \bibnamefont{Weston}},
  \bibinfo{author}{\bibfnamefont{H.~M.} \bibnamefont{Chrzanowski}},
  \bibinfo{author}{\bibfnamefont{L.~K.} \bibnamefont{Shalm}},
  \bibinfo{author}{\bibfnamefont{V.~B.} \bibnamefont{Verma}},
  \bibinfo{author}{\bibfnamefont{S.~W.} \bibnamefont{Nam}}, \bibnamefont{and}
  \bibinfo{author}{\bibfnamefont{G.~J.} \bibnamefont{Pryde}}
  (\bibinfo{year}{2017}), \eprint{arXiv:1707.08977 [quant-ph]}.

\bibitem[{\citenamefont{Higgins et~al.}(2007)\citenamefont{Higgins, Berry,
  Bartlett, Wiseman, and Pryde}}]{higgins2007}
\bibinfo{author}{\bibfnamefont{B.~L.} \bibnamefont{Higgins}},
  \bibinfo{author}{\bibfnamefont{D.~W.} \bibnamefont{Berry}},
  \bibinfo{author}{\bibfnamefont{S.~D.} \bibnamefont{Bartlett}},
  \bibinfo{author}{\bibfnamefont{H.~M.} \bibnamefont{Wiseman}},
  \bibnamefont{and} \bibinfo{author}{\bibfnamefont{G.~J.} \bibnamefont{Pryde}},
  \bibinfo{journal}{Nature} \textbf{\bibinfo{volume}{450}},
  \bibinfo{pages}{393} (\bibinfo{year}{2007}).

\bibitem[{\citenamefont{Knysh et~al.}(2011)\citenamefont{Knysh, Smelyanskiy,
  and Durkin}}]{knysh2011}
\bibinfo{author}{\bibfnamefont{S.}~\bibnamefont{Knysh}},
  \bibinfo{author}{\bibfnamefont{V.~N.} \bibnamefont{Smelyanskiy}},
  \bibnamefont{and} \bibinfo{author}{\bibfnamefont{G.~A.}
  \bibnamefont{Durkin}}, \bibinfo{journal}{Phys. Rev. A}
  \textbf{\bibinfo{volume}{83}}, \bibinfo{pages}{021804}
  (\bibinfo{year}{2011}).

\bibitem[{\citenamefont{Escher et~al.}(2011)\citenamefont{Escher,
  de~Matos~Filho, and Davidovich}}]{escher2011}
\bibinfo{author}{\bibfnamefont{B.~M.} \bibnamefont{Escher}},
  \bibinfo{author}{\bibfnamefont{R.~L.} \bibnamefont{de~Matos~Filho}},
  \bibnamefont{and}
  \bibinfo{author}{\bibfnamefont{L.}~\bibnamefont{Davidovich}},
  \bibinfo{journal}{Nature Phys.} \textbf{\bibinfo{volume}{7}},
  \bibinfo{pages}{406} (\bibinfo{year}{2011}).

\bibitem[{\citenamefont{Jarzyna and
  Demkowicz-Dobrza{\'n}ski}(2012)}]{jarzyna2012}
\bibinfo{author}{\bibfnamefont{M.}~\bibnamefont{Jarzyna}} \bibnamefont{and}
  \bibinfo{author}{\bibfnamefont{R.}~\bibnamefont{Demkowicz-Dobrza{\'n}ski}},
  \bibinfo{journal}{Phys. Rev. A} \textbf{\bibinfo{volume}{85}},
  \bibinfo{pages}{011801} (\bibinfo{year}{2012}).

\bibitem[{\citenamefont{Taylor and Bowen}(2016)}]{taylor2016}
\bibinfo{author}{\bibfnamefont{M.~A.} \bibnamefont{Taylor}} \bibnamefont{and}
  \bibinfo{author}{\bibfnamefont{W.~P.} \bibnamefont{Bowen}},
  \bibinfo{journal}{Phys. Rep.} \textbf{\bibinfo{volume}{615}},
  \bibinfo{pages}{1} (\bibinfo{year}{2016}).

\bibitem[{\citenamefont{Juffmann et~al.}(2016)\citenamefont{Juffmann, Klopfer,
  Frankort, Haslinger, and Kasevich}}]{juffmann2016}
\bibinfo{author}{\bibfnamefont{T.}~\bibnamefont{Juffmann}},
  \bibinfo{author}{\bibfnamefont{B.~B.} \bibnamefont{Klopfer}},
  \bibinfo{author}{\bibfnamefont{T.~L.~I.} \bibnamefont{Frankort}},
  \bibinfo{author}{\bibfnamefont{P.}~\bibnamefont{Haslinger}},
  \bibnamefont{and} \bibinfo{author}{\bibfnamefont{M.~A.}
  \bibnamefont{Kasevich}}, \bibinfo{journal}{Nat. Commun.}
  \textbf{\bibinfo{volume}{7:12858}} (\bibinfo{year}{2016}).

\bibitem[{\citenamefont{Juffmann et~al.}(2017)\citenamefont{Juffmann, Koppell,
  Klopfer, Ophus, Glaeser, and Kasevich}}]{juffmann2017}
\bibinfo{author}{\bibfnamefont{T.}~\bibnamefont{Juffmann}},
  \bibinfo{author}{\bibfnamefont{S.~A.} \bibnamefont{Koppell}},
  \bibinfo{author}{\bibfnamefont{B.~B.} \bibnamefont{Klopfer}},
  \bibinfo{author}{\bibfnamefont{C.}~\bibnamefont{Ophus}},
  \bibinfo{author}{\bibfnamefont{R.~M.} \bibnamefont{Glaeser}},
  \bibnamefont{and} \bibinfo{author}{\bibfnamefont{M.~A.}
  \bibnamefont{Kasevich}}, \bibinfo{journal}{Scientific Reports}
  \textbf{\bibinfo{volume}{7:1699}} (\bibinfo{year}{2017}).

\bibitem[{\citenamefont{Dorner et~al.}(2009)\citenamefont{Dorner,
  Demkowicz-Dobrzanski, Smith, Lundeen, Wasilewski, Banaszek, and
  Walmsley}}]{dorner2009}
\bibinfo{author}{\bibfnamefont{U.}~\bibnamefont{Dorner}},
  \bibinfo{author}{\bibfnamefont{R.}~\bibnamefont{Demkowicz-Dobrzanski}},
  \bibinfo{author}{\bibfnamefont{B.~J.} \bibnamefont{Smith}},
  \bibinfo{author}{\bibfnamefont{J.~S.} \bibnamefont{Lundeen}},
  \bibinfo{author}{\bibfnamefont{W.}~\bibnamefont{Wasilewski}},
  \bibinfo{author}{\bibfnamefont{K.}~\bibnamefont{Banaszek}}, \bibnamefont{and}
  \bibinfo{author}{\bibfnamefont{I.~A.} \bibnamefont{Walmsley}},
  \bibinfo{journal}{Phys. Rev. Lett.} \textbf{\bibinfo{volume}{102}},
  \bibinfo{pages}{040403} (\bibinfo{year}{2009}).

\bibitem[{\citenamefont{Demkowicz-Dobrza{\'n}ski}(2010)}]{demkowicz2010}
\bibinfo{author}{\bibfnamefont{R.}~\bibnamefont{Demkowicz-Dobrza{\'n}ski}},
  \bibinfo{journal}{Laser Phys.} \textbf{\bibinfo{volume}{20}},
  \bibinfo{pages}{1197} (\bibinfo{year}{2010}).

\bibitem[{\citenamefont{Demkowicz-Dobrza{\'n}ski and
  Maccone}(2014)}]{demkowicz2014entangle}
\bibinfo{author}{\bibfnamefont{R.}~\bibnamefont{Demkowicz-Dobrza{\'n}ski}}
  \bibnamefont{and} \bibinfo{author}{\bibfnamefont{L.}~\bibnamefont{Maccone}},
  \bibinfo{journal}{Phys. Rev. Lett.} \textbf{\bibinfo{volume}{113}},
  \bibinfo{pages}{250801} (\bibinfo{year}{2014}).

\bibitem[{\citenamefont{Helstrom}(1976)}]{helstrom1976}
\bibinfo{author}{\bibfnamefont{C.~W.} \bibnamefont{Helstrom}},
  \emph{\bibinfo{title}{Quantum detection and estimation theory}}
  (\bibinfo{publisher}{Academic press}, \bibinfo{year}{1976}).

\bibitem[{\citenamefont{Braunstein and Caves}(1994)}]{braunstein1994}
\bibinfo{author}{\bibfnamefont{S.~L.} \bibnamefont{Braunstein}}
  \bibnamefont{and} \bibinfo{author}{\bibfnamefont{C.~M.} \bibnamefont{Caves}},
  \bibinfo{journal}{Phys. Rev. Lett.} \textbf{\bibinfo{volume}{72}},
  \bibinfo{pages}{3439} (\bibinfo{year}{1994}).

\bibitem[{\citenamefont{Van~der Vaart}(2000)}]{van2000}
\bibinfo{author}{\bibfnamefont{A.~W.} \bibnamefont{Van~der Vaart}},
  \emph{\bibinfo{title}{Asymptotic statistics}}, vol.~\bibinfo{volume}{3}
  (\bibinfo{publisher}{Cambridge university press}, \bibinfo{year}{2000}).

\bibitem[{\citenamefont{Demkowicz-Dobrzanski
  et~al.}(2009)\citenamefont{Demkowicz-Dobrzanski, Dorner, Smith, Lundeen,
  Wasilewski, Banaszek, and Walmsley}}]{demkowicz2009}
\bibinfo{author}{\bibfnamefont{R.}~\bibnamefont{Demkowicz-Dobrzanski}},
  \bibinfo{author}{\bibfnamefont{U.}~\bibnamefont{Dorner}},
  \bibinfo{author}{\bibfnamefont{B.~J.} \bibnamefont{Smith}},
  \bibinfo{author}{\bibfnamefont{J.~S.} \bibnamefont{Lundeen}},
  \bibinfo{author}{\bibfnamefont{W.}~\bibnamefont{Wasilewski}},
  \bibinfo{author}{\bibfnamefont{K.}~\bibnamefont{Banaszek}}, \bibnamefont{and}
  \bibinfo{author}{\bibfnamefont{I.~A.} \bibnamefont{Walmsley}},
  \bibinfo{journal}{Phys. Rev. A} \textbf{\bibinfo{volume}{80}},
  \bibinfo{pages}{013825} (\bibinfo{year}{2009}).

\bibitem[{\citenamefont{Corless et~al.}(1996)\citenamefont{Corless, Gonnet,
  Hare, Jeffrey, and Knuth}}]{corless1996}
\bibinfo{author}{\bibfnamefont{R.~M.} \bibnamefont{Corless}},
  \bibinfo{author}{\bibfnamefont{G.~H.} \bibnamefont{Gonnet}},
  \bibinfo{author}{\bibfnamefont{D.~E.} \bibnamefont{Hare}},
  \bibinfo{author}{\bibfnamefont{D.~J.} \bibnamefont{Jeffrey}},
  \bibnamefont{and} \bibinfo{author}{\bibfnamefont{D.~E.} \bibnamefont{Knuth}},
  \bibinfo{journal}{Adv. Comput. Math.} \textbf{\bibinfo{volume}{5}},
  \bibinfo{pages}{329} (\bibinfo{year}{1996}).

\bibitem[{\citenamefont{Demkowicz-Dobrza{\'n}ski
  et~al.}(2013)\citenamefont{Demkowicz-Dobrza{\'n}ski, Banaszek, and
  Schnabel}}]{demkowicz2013}
\bibinfo{author}{\bibfnamefont{R.}~\bibnamefont{Demkowicz-Dobrza{\'n}ski}},
  \bibinfo{author}{\bibfnamefont{K.}~\bibnamefont{Banaszek}}, \bibnamefont{and}
  \bibinfo{author}{\bibfnamefont{R.}~\bibnamefont{Schnabel}},
  \bibinfo{journal}{Phys. Rev. A} \textbf{\bibinfo{volume}{88}},
  \bibinfo{pages}{041802} (\bibinfo{year}{2013}).

\bibitem[{\citenamefont{Aspachs et~al.}(2009)\citenamefont{Aspachs,
  Calsamiglia, Munoz-Tapia, and Bagan}}]{aspachs2009}
\bibinfo{author}{\bibfnamefont{M.}~\bibnamefont{Aspachs}},
  \bibinfo{author}{\bibfnamefont{J.}~\bibnamefont{Calsamiglia}},
  \bibinfo{author}{\bibfnamefont{R.}~\bibnamefont{Munoz-Tapia}},
  \bibnamefont{and} \bibinfo{author}{\bibfnamefont{E.}~\bibnamefont{Bagan}},
  \bibinfo{journal}{Phys. Rev. A} \textbf{\bibinfo{volume}{79}},
  \bibinfo{pages}{033834} (\bibinfo{year}{2009}).

\bibitem[{\citenamefont{Berry et~al.}(2009)\citenamefont{Berry, Higgins,
  Bartlett, Mitchell, Pryde, and Wiseman}}]{berry2009}
\bibinfo{author}{\bibfnamefont{D.~W.} \bibnamefont{Berry}},
  \bibinfo{author}{\bibfnamefont{B.~L.} \bibnamefont{Higgins}},
  \bibinfo{author}{\bibfnamefont{S.~D.} \bibnamefont{Bartlett}},
  \bibinfo{author}{\bibfnamefont{M.~W.} \bibnamefont{Mitchell}},
  \bibinfo{author}{\bibfnamefont{G.~J.} \bibnamefont{Pryde}}, \bibnamefont{and}
  \bibinfo{author}{\bibfnamefont{H.~M.} \bibnamefont{Wiseman}},
  \bibinfo{journal}{Phys. Rev. A} \textbf{\bibinfo{volume}{80}},
  \bibinfo{pages}{052114} (\bibinfo{year}{2009}).

\bibitem[{\citenamefont{Kacprowicz et~al.}(2010)\citenamefont{Kacprowicz,
  Demkowicz-Dobrza{\'n}ski, Wasilewski, Banaszek, and
  Walmsley}}]{kacprowicz2010}
\bibinfo{author}{\bibfnamefont{M.}~\bibnamefont{Kacprowicz}},
  \bibinfo{author}{\bibfnamefont{R.}~\bibnamefont{Demkowicz-Dobrza{\'n}ski}},
  \bibinfo{author}{\bibfnamefont{W.}~\bibnamefont{Wasilewski}},
  \bibinfo{author}{\bibfnamefont{K.}~\bibnamefont{Banaszek}}, \bibnamefont{and}
  \bibinfo{author}{\bibfnamefont{I.~A.} \bibnamefont{Walmsley}},
  \bibinfo{journal}{Nature Photon.} \textbf{\bibinfo{volume}{4}},
  \bibinfo{pages}{357} (\bibinfo{year}{2010}).

\bibitem[{\citenamefont{Thomas-Peter et~al.}(2011)\citenamefont{Thomas-Peter,
  Smith, Datta, Zhang, Dorner, and Walmsley}}]{thomas2011}
\bibinfo{author}{\bibfnamefont{N.}~\bibnamefont{Thomas-Peter}},
  \bibinfo{author}{\bibfnamefont{B.~J.} \bibnamefont{Smith}},
  \bibinfo{author}{\bibfnamefont{A.}~\bibnamefont{Datta}},
  \bibinfo{author}{\bibfnamefont{L.}~\bibnamefont{Zhang}},
  \bibinfo{author}{\bibfnamefont{U.}~\bibnamefont{Dorner}}, \bibnamefont{and}
  \bibinfo{author}{\bibfnamefont{I.~A.} \bibnamefont{Walmsley}},
  \bibinfo{journal}{Phys. Rev. Lett.} \textbf{\bibinfo{volume}{107}},
  \bibinfo{pages}{113603} (\bibinfo{year}{2011}).

\bibitem[{\citenamefont{Cable and Durkin}(2010)}]{cable2010}
\bibinfo{author}{\bibfnamefont{H.}~\bibnamefont{Cable}} \bibnamefont{and}
  \bibinfo{author}{\bibfnamefont{G.~A.} \bibnamefont{Durkin}},
  \bibinfo{journal}{Phys. Rev. Lett.} \textbf{\bibinfo{volume}{105}},
  \bibinfo{pages}{013603} (\bibinfo{year}{2010}).

\bibitem[{\citenamefont{Collaboration et~al.}(2011)}]{ligo2011}
\bibinfo{author}{\bibfnamefont{L.~S.} \bibnamefont{Collaboration}}
  \bibnamefont{et~al.}, \bibinfo{journal}{Nature Phys.}
  \textbf{\bibinfo{volume}{7}}, \bibinfo{pages}{962} (\bibinfo{year}{2011}).

\bibitem[{\citenamefont{Grote et~al.}(2013)\citenamefont{Grote, Danzmann,
  Dooley, Schnabel, Slutsky, and Vahlbruch}}]{grote2013}
\bibinfo{author}{\bibfnamefont{H.}~\bibnamefont{Grote}},
  \bibinfo{author}{\bibfnamefont{K.}~\bibnamefont{Danzmann}},
  \bibinfo{author}{\bibfnamefont{K.~L.} \bibnamefont{Dooley}},
  \bibinfo{author}{\bibfnamefont{R.}~\bibnamefont{Schnabel}},
  \bibinfo{author}{\bibfnamefont{J.}~\bibnamefont{Slutsky}}, \bibnamefont{and}
  \bibinfo{author}{\bibfnamefont{H.}~\bibnamefont{Vahlbruch}},
  \bibinfo{journal}{Phys. Rev. Lett.} \textbf{\bibinfo{volume}{110}},
  \bibinfo{pages}{181101} (\bibinfo{year}{2013}).

\bibitem[{\citenamefont{Aasi et~al.}(2013)\citenamefont{Aasi, Abadie, Abbott,
  Abbott, Abbott, Abernathy, Adams, Adams, Addesso, Adhikari
  et~al.}}]{aasi2013}
\bibinfo{author}{\bibfnamefont{J.}~\bibnamefont{Aasi}},
  \bibinfo{author}{\bibfnamefont{J.}~\bibnamefont{Abadie}},
  \bibinfo{author}{\bibfnamefont{B.~P.} \bibnamefont{Abbott}},
  \bibinfo{author}{\bibfnamefont{R.}~\bibnamefont{Abbott}},
  \bibinfo{author}{\bibfnamefont{T.~D.} \bibnamefont{Abbott}},
  \bibinfo{author}{\bibfnamefont{M.~R.} \bibnamefont{Abernathy}},
  \bibinfo{author}{\bibfnamefont{C.}~\bibnamefont{Adams}},
  \bibinfo{author}{\bibfnamefont{T.}~\bibnamefont{Adams}},
  \bibinfo{author}{\bibfnamefont{P.}~\bibnamefont{Addesso}},
  \bibinfo{author}{\bibfnamefont{R.~X.} \bibnamefont{Adhikari}},
  \bibnamefont{et~al.}, \bibinfo{journal}{Nature Photon.}
  \textbf{\bibinfo{volume}{7}}, \bibinfo{pages}{613} (\bibinfo{year}{2013}).

\bibitem[{\citenamefont{Loudon and Knight}(1987)}]{loudon1987}
\bibinfo{author}{\bibfnamefont{R.}~\bibnamefont{Loudon}} \bibnamefont{and}
  \bibinfo{author}{\bibfnamefont{P.~L.} \bibnamefont{Knight}},
  \bibinfo{journal}{J. Mod. Opt.} \textbf{\bibinfo{volume}{34}},
  \bibinfo{pages}{709} (\bibinfo{year}{1987}).

\bibitem[{\citenamefont{Pitkin et~al.}(2011)\citenamefont{Pitkin, Reid, Rowan,
  and Hough}}]{pitkin2011}
\bibinfo{author}{\bibfnamefont{M.}~\bibnamefont{Pitkin}},
  \bibinfo{author}{\bibfnamefont{S.}~\bibnamefont{Reid}},
  \bibinfo{author}{\bibfnamefont{S.}~\bibnamefont{Rowan}}, \bibnamefont{and}
  \bibinfo{author}{\bibfnamefont{J.}~\bibnamefont{Hough}},
  \bibinfo{journal}{Living Reviews in Relativity}
  \textbf{\bibinfo{volume}{14}}, \bibinfo{pages}{5} (\bibinfo{year}{2011}).

\bibitem[{\citenamefont{Pinel et~al.}(2013)\citenamefont{Pinel, Jian, Treps,
  Fabre, and Braun}}]{pinel2013}
\bibinfo{author}{\bibfnamefont{O.}~\bibnamefont{Pinel}},
  \bibinfo{author}{\bibfnamefont{P.}~\bibnamefont{Jian}},
  \bibinfo{author}{\bibfnamefont{N.}~\bibnamefont{Treps}},
  \bibinfo{author}{\bibfnamefont{C.}~\bibnamefont{Fabre}}, \bibnamefont{and}
  \bibinfo{author}{\bibfnamefont{D.}~\bibnamefont{Braun}},
  \bibinfo{journal}{Phys. Rev. A} \textbf{\bibinfo{volume}{88}},
  \bibinfo{pages}{040102} (\bibinfo{year}{2013}).

\bibitem[{\citenamefont{Grimmett and Stirzaker}(2001)}]{grimmett2001}
\bibinfo{author}{\bibfnamefont{G.}~\bibnamefont{Grimmett}} \bibnamefont{and}
  \bibinfo{author}{\bibfnamefont{D.}~\bibnamefont{Stirzaker}},
  \emph{\bibinfo{title}{Probability and random processes}}
  (\bibinfo{publisher}{Oxford university press}, \bibinfo{year}{2001}).

\end{thebibliography}

%%%%%%%%%% APPENDIX

%%%%%%%%%% APPENDIX

\appendix
\section{Multipass quantum limits}
Here we discuss the multi-pass versions of the bounds on QFI. As stated in the main text: the upper bound on QFI per lost photon for TM phase estimation using any input state is \cite{escher2011}:
\begin{equation}
 Q_{\text{TM}}\equiv\tr(\hat N\varrho^\text{in})\eta\big/(1-\eta) \geq F(\varrho^{\text{out}}).
\end{equation}
In the derivation of this equation $\eta$ refers to the total transmissivity that the probe state experiences and bounds the QFI associated with the total phase shift experienced by the probe state. In the multi-pass case the transmissivitiy is instead $\eta^k$ and so $\eta^k$ replaces all occurrences of $\eta$ to adapt this equation to the multi-pass case. In addition to the reduced transmissivity, the total phase shift is also magnified by a factor of $k$. The RMSE of estimates of this total phase shift $k\theta$ are bound by the limit on QFI  with the reduced transmissivity: $\tr(\hat N\varrho^\text{in})\eta^k\big/(1-\eta^k)$. This implies a limit on the estimation precision of $\theta$ by the following argument: since $\Delta^2 (k \theta) = (1/k^2) \Delta^2 \theta$, then according to the Cram\'{e}r-Rao bound the QFI associated with estimating $\theta$ must be a factor of $k^2$ larger than the QFI associated with estimating $k\theta$. As such $Q_{\text{TM}}(k)=k^2\tr(\hat N\varrho^\text{in})\eta^k\big/(1-\eta^k)$. Finally, dividing by the number of lost photons gives: $Q^\prime_{\text{TM}}(k)=k^2 \eta^k\big/(1-\eta^k)^2$. $Q_{\text{SM}}(k)$ can be found in the same way. These equations could be derived more formally by considering how the QFI transforms when multi-pass setups are used as is done in the main text for QFI per lost photon.

\section{Details for Gaussian strategies}
\begin{figure}[h]
 \includegraphics[width=0.6\columnwidth]{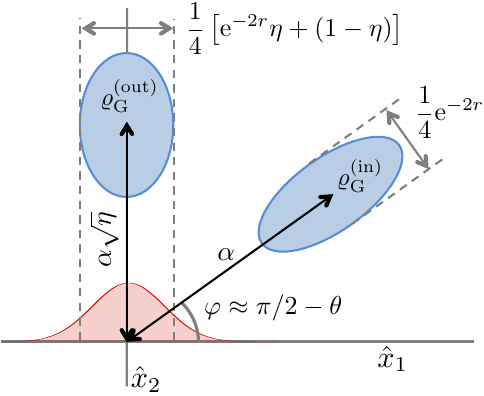}
 \caption{Depiction of the covariances of the phase-squeezed Gaussian state discussed in the text. The homodyne signal is depicted along the horizontal axis.}
 \label{gauss_def}
\end{figure}

Here we identify a Gaussian input state which maximizes the QFI related to the phase dependence of the displacement vector \cite{pinel2013}, noting that QFI related to the phase dependence of the $\mathbf{\Gamma}$ is not retrieved by homodyne measurements. Specifically it was found in Ref.~\cite{aspachs2009} that homodyne measurements on squeezed thermal states are far from optimal.
Following the definitions in the main text, the Gaussian input state $\varrho^{\text{in}}_{\text{G}}$ has a displacement vector $\bm{d}^{\text{in}}=R(\varphi)(\alpha,0)^{T}$ and a
covariance matrix:
\begin{equation}
\mathbf{\Gamma}^{\text{in}}=\frac{1}{4}R(\varphi)
\begin{pmatrix}
\e^{2r} & 0 \\
0 & \e^{-2r}
\end{pmatrix}
R^{T}(\varphi)
\end{equation}
where
$R(\varphi) = \begin{pmatrix}\cos\varphi &-\sin\varphi \\ \sin\varphi & \cos\varphi \end{pmatrix}
$
 is the usual rotation matrix. Following the lossy phase $\Lambda(\eta,\theta)$ the displacement vector of the output state,  $\varrho^{\text{out}}_{\text{G}}$, is $\bm{d}^{\text{out}}=R(\varphi+\theta)(\sqrt{\eta}\alpha,0)^{T}$ and the covariance matrix is:
\begin{equation}
\mathbf{\Gamma}^{\text{out}}=\frac{1}{4}R(\varphi+\theta)
\begin{pmatrix}
\e^{2r}\eta \!+\! 1\!-\!\eta & 0 \\
0 & \e^{-2r}\eta \!+\! 1\!-\!\eta
\end{pmatrix}
R^{T}(\varphi+\theta).
\end{equation}
Loss introduces a factor of $\eta$ into each covariance element and adds $(1-\eta)$ to the diagonal elements \cite{aspachs2009}. This can be thought of as introducing some vacuum noise from the environment mode (the reader should note that we use conventions to define Gaussian states and the covariance matrix in accordance with Ref.\cite{loudon1987}).
The states $\varrho^{\text{in}}_{\text{G}}$ and $\varrho^{\text{out}}_{\text{G}}$ are depicted in Fig.~\ref{gauss_def}.
The Fisher information from a normally distributed measurement outcome $x$ with a mean of $\mu$ and a variance of $V$ about a parameter $\theta$ can be found using the definition of the Fisher information $\int\left[\frac{\partial}{\partial \theta}\ln p(x|\theta)\right]^2 p(x|\theta)dx$ \cite{van2000}. This gives $F_{x} = {\dot \mu}^2/V+ {\dot V}^2/(2V^2)$, where dots indicate derivatives with respect to the parameter to be estimated, such that we can insert the mean and variance of the homodyne signal upon measurement of $\varrho^{\text{out}}_{\text{G}}$, given by $d^{\text{out}}_{1}$ and $\Gamma^{\text{out}}_{11}$. We maximize the term in the Fisher information involving phase dependence of the displacement vector $({\dot d}^{\text{out}}_{1})^2/\Gamma^{\text{out}}_{11}$ by setting $\varphi = \pi/2-\theta$ to obtain:
\begin{equation}
\begin{split}
F_{\hat x}(\varrho_{\theta})\big|_{\varphi = \pi/2 - \theta} &=  \frac{4 \alpha^2 \eta}{1 + ( e^{-2 r}-1) \eta}\\
&= \frac{4 \eta  \left(n-n_{\text{sq}}\right)}{2 \eta  n_{\text{sq}}-2 \eta  \sqrt{n_{\text{sq}} (n_{\text{sq}}+1)}+1}.
\end{split}
\end{equation}
For a given mean number of photons, $n$, this has a maximum value when:
\begin{equation}
n_{\text{sq}}=-\frac{\left(\sqrt{1-4 (\eta -1) \eta  n}-1\right)^2}{4 (\eta -1) \left(\sqrt{1-4 (\eta -1) \eta  n}-\eta \right)}
\end{equation}
and a Fisher information of:
\begin{equation}
2\eta\frac{ 1+2n(1-\eta)-\sqrt{1+4n(1-\eta)\eta}}{(\eta-1)^2}.
\end{equation}
This expression could be used to compare this scheme to other schemes which are designed to operate in the extremely low photon number regime.

\section{More general interferometers}
We consider the interrogation of a SM lossy phase embedded in a setup, such that between each interrogation interference occurs between the outgoing mode from the phase and another optical mode. Specifically we investigate an optical setup in which the coherent-state field amplitudes in two modes, $(\alpha,\beta)^{T}$ of the state $|\alpha\rangle|\beta\rangle$, are transformed by the matrix:
\begin{equation} \label{general_network}
T = \left[
\begin{pmatrix}
1 & 0\\
0 & \e^{\ti k \theta+ \phi}\eta^{k/2}
\end{pmatrix}
\begin{pmatrix}
\cos(\xi/2) & -\sin(\xi/2)\\
\sin(\xi/2) & \cos(\xi/2)
\end{pmatrix}
\right]^h
\end{equation}
where the right-hand matrix is an arbitrary-reflectivity beam-splitter and the left-hand matrix is the application of a control phase $\phi$ plus the phase and damping associated with the lossy phase $\Lambda^k(\eta,\theta)$. An optical setup which can implement this is shown in Fig.~3 of the main text. We performed a heuristic optimization of the QFI per lost photon produced by this strategy using the input state $|\alpha\rangle|0\rangle$ for a series of values of $\eta = 0.1,0.2,...,0.9$ over the parameters $h, k,\xi$ and $\varphi=k\theta+\phi$ whilst constraining $k$ and $h$ to be integers. In order to remove the effects of restricting to discrete $k$, we performed an optimization of $\ln^2\eta\, \mathcal{F}^\prime(\varrho)$, which we previously found to be of the form $f(\varphi,\gamma = \eta^k)$, and varied ($h,\varphi, \xi$) and  $\gamma$. This yielded a classical strategy for which $\mathcal{F}^\prime(\varrho_{\text{Cl}}) =4\times 0.94/\ln^2\eta$. The reduction in RMSE per incident photon, afforded by non-classical techniques, was found to be $3.6\%$ on average for $\eta = 0.1,0.2,...,0.9$ and $3.0\%$ when $k$ was permitted to be non-integer.

\section{Optimization details}
Given the two-mode coherent state input: $|\psi^{\text{in}}_\text{Cl}\rangle=|\alpha_1^{\text{in}}\rangle|0\rangle$, where $\alpha_1^{\text{in}} \in \mathbb{R}_{> 0}$ without loss of generality, the output state after the network described by the transfer matrix $T$, given in Eq.~(\ref{general_network}), will be a two-mode coherent state $|\psi^{\text{out}}_\text{Cl}\rangle=|\alpha_1^{\text{in}}T_{11}(\theta)\rangle|\alpha_1^{\text{in}}T_{21}(\theta)\rangle$ which has a QFI of:
\begin{equation}
4\left(\alpha_1^{\text{in}}\right)^2\left(\left|\frac{\partial  T_{11}}{\partial \theta}\right|^2+\left|\frac{\partial  T_{21}}{\partial \theta}\right|^2\right).
\end{equation}
where $T_{11}$ and $T_{21}$ are matrix elements in the first column of $T$. This can be seen from the results in Ref.~\cite{pinel2013} and using the fact that QFI of a system is additive over separable sub-systems. Alternatively this can be computed directly from the definition of the QFI. Therefore, the QFI per lost photon of $|\psi^{\text{out}}_\text{Cl}\rangle$ is
\begin{equation}
4\frac{\left|\frac{\partial  T_{11}}{\partial \theta}\right|^2+\left|\frac{\partial  T_{21}}{\partial \theta}\right|^2}{1-|T_{11}|^2-|T_{21}|^2}.
\end{equation}
This expression was heuristically optimized. Through preliminary optimizations it was noticed that $\varphi = 0$ and $k=1$ were generally the best values for these parameters. Then we numerically optimized over $\xi \in(0.1/h,6/h)$ and $h\in \mathbb{N}=\{1,2,\ldots\}$ whilst fixing $\phi = 0$ and $k=1$ for each value of $\eta\in\{0.1,0.2,\ldots0.9\}$ using a standard numerical optimization package.  $\xi$ was restricted to be greater than $0.1/h$ to improve numerical stability. The results of these optimizations are shown in Table \ref{opt_table}.

\begin{center}
\begin{table}[h]
\setlength{\tabcolsep}{13pt}
\renewcommand{\arraystretch}{1.5}
  \begin{tabular}{ c c c c}
    \hline \hline
    $\eta$ (set) & $h$ (varied) & $h\xi$ (varied) & $\mathcal{F}^{\prime}/ Q^{\prime}_{\text{SM}}$ \\ \hline
    0.1 & 1 & NA &0.900000 \\
    0.2 & 2 & 0.0100001& 0.927824 \\
0.3 & 3 & 0.0100005&0.922368 \\
0.4 & 3 &0.0100004&0.924007 \\
0.5 & 5 & 0.0100000&0.930968 \\
0.6 & 6 & 0.0100009& 0.933002 \\
0.7 & 9 & 0.0100000& 0.934944 \\
0.8 & 15 & 0.0100001& 0.935805 \\
0.9 & 32 & 0.0100076&0.936049 \\
    \hline \hline
  \end{tabular}
\caption{Optimized values of $\mathcal{F}^{\prime}$ for a set of $\eta$. The values quoted have been rounded to six significant figures where appropriate. }
\label{opt_table}
\end{table}
\end{center}
The results shown in Table \ref{opt_table} appear to indicate that the best strategy always has a small $h\xi$ independent of $\eta$. However, as stated in Appendix C it is necessary to address the effects of confining this strategy to discrete numbers of interrogations. We performed a numerical optimization of $\ln^2\eta\, F^\prime(\varrho)$, which we previously found to be of the form $f(\varphi,\gamma = \eta^k)$, by varying the optical setup ($h,\xi$) and $\gamma$. $\varphi$ was again set to zero. The result of this optimization is shown in Table \ref{cont_opt}. This result indicates that when $\eta$ is close to one, a strategy which has a small $h\xi$ is not optimal.
\begin{center}
\begin{table}[h]
\setlength{\tabcolsep}{6pt}
\renewcommand{\arraystretch}{1.5}
  \begin{tabular}{ c c c c}
    \hline \hline
    $\gamma$ (varied) & $h$ (varied) & $h\xi$ (varied) & $\mathcal{F}^{\prime}/ \lim_{k\rightarrow 0}Q^{\prime}_{\text{SM}}(k)$ \\ \hline
   0.988747 & 510 & 5.12423 &0.940333 \\

    \hline \hline
  \end{tabular}
\caption{Optimized values of $\mathcal{F}^{\prime}$ normalized to the quantum limit  $\lim_{k\rightarrow 0}Q_{\text{SM}}^{\prime}(k)$. The values quoted have been rounded to six significant figures where appropriate. }
\label{cont_opt}
\end{table}
\end{center}

\section{General networks for two-mode phase estimation}
We were unable to find schemes for TM phase estimation which went above the optimal classical MP $\mathcal{F}^\prime$, given in Eq.~(4) of the main text, without using any reference beams. However, we note that the phase difference between the two modes $\theta_\delta=\theta_1-\theta_2$, with equally lossy and uncorrelated phases in each mode $\theta_1$ and $\theta_2$, could be estimated by separately interrogating the phase in each arm, and utilizing two additional modes to perform the estimation following the SM strategies.

Measuring each phase, $\theta_1$ and $\theta_2$, equally precisely will require twice as many photons to be lost in comparison with the estimation of a single phase to the same precision. The RMSE of the phase difference will be twice as big as it is for each phase $\theta_1$ and $\theta_2$ since $\Delta^2\theta_\delta=\Delta^2\theta_1+\Delta^2\theta_2$ \cite{grimmett2001}. Therefore, this procedure would provide one quarter of the Fisher information per lost photon about the phase difference as in the case of SM phase estimation. Hence the ratio between optimal quantum and this classical strategy for TM phase estimation, when reference beams are used, would be as for SM phase estimation in the main text.

\section{Experimental imperfections}
To understand how deviations from the ideal case affect the scope for quantum enhancement in a multi-pass setup we consider optical loss $1-\eta_{{p}}$ encountered before incidence upon the lossy-phase $\Lambda(\eta,\theta)$, round-trip loss $1-\eta_{{r}}$ encountered between applications of $\Lambda(\eta,\theta)$ and measurement loss $1-\eta_{m}$ encountered after the last application of $\Lambda(\eta,\theta)$. Thus, for $k$ passes the total transmissivity is $\eta_{\text{tot}}=\eta^k\eta_{p}\eta_{m}\eta_{r}^{k-1}$ and the number of photons lost via the lossy-phase is:
\begin{equation}
\begin{split}
 \langle\hat{n}_{l}\rangle&= \sum_{i=1}^k (1-\eta)\eta_p\tr(\hat N \varrho^{\text{in}})\eta_r^{i-1}\eta^{i-1}  \\
 &=\tr(\hat{N}\varrho^{\text{in}}) \eta_p(1-\eta)(1-\eta^k\eta_{r}^k)\big/(1-\eta\eta_{r})
\end{split}
\end{equation}
which has been found by summing over the number of scattered photons each time the state is incident upon the lossy-phase. With these additional losses, we consider the resource to be the QFI per photon lost via the lossy phase (which is equivalent to considering the QFI per photon incident on the lossy phase since the number lost via the phase is proportional to the number incident upon the phase).

We consider the capabilities of an optimal quantum TM MP strategy by using the saturable bounds on QFI for any input state:
$$
k^2 \tr(\hat N\varrho^\text{in})\eta_\text{tot}\big/(1-\eta_\text{tot})
$$
and divide through by $\langle\hat{n}_{l}\rangle$ to find the highest QFI per lost photon for any input state. This equation has been obtained from Eq.~(\ref{twomodebound}) following the steps described in Appendix A.
Similarly, we use the QFI for a classical TM SP strategy $\tr(\hat{N}\varrho^\text{in})\eta$\, \cite{demkowicz2010} to
 obtain the QFI per lost photon for a classical TM MP strategy as
 $
 \tr(\hat{N}\varrho^\text{in})\eta_\text{tot}k^2/ \langle\hat{n}_{l}\rangle.
$

When these additional losses are included we were unable to find general closed form expressions for the optimal number of passes for either classical or optimal quantum MP strategies. Because of this we could not compute expressions for optimal QFI using optimal quantum or classical states with imperfect multi-pass strategies. However, using numerical methods, we were able to clarify situations for which quantum strategies grant precision gains which are larger than $20\%$ (i.e. larger than in the ideal case). We focused on evaluating the difference in achievable precision between a quantum strategy and an equivalent classical strategy, i.e. with the same losses. It should be noted that the advantage granted here by quantum techniques is not fundamental, in the sense that it could only be achieved by non-classical techniques, since better components could improve the classical strategy.

For a given input state the QFI obtainable from a multi-pass scheme is dependent on the total phase shift applied and on the overall transmissivity, both of which can be found from $k,\eta_p\eta_m,\eta$ and $\eta_r$. The number of photons lost via the phase is a function of the form $\eta_p\times \tr(\hat{N}\varrho^{\text{in}})\times q(k,\eta,\eta_r)$ where $q(k,\eta,\eta_r)=(1-\eta)(1-\eta^k\eta_{r}^k)\big/(1-\eta\eta_{r})$. Therefore, in the case that the QFI for a given input state is proportional to the number of photons it contains, as is the case for both the optimal quantum state and the optimal classical state, the QFI per photon lost via the phase is a function of the form $(1/\eta_p)\times z(\eta_p\eta_m,k,\eta,\eta_r)$. Therefore, the ratio between the achievable precisions using quantum and classical states is independent of $\eta_p$ once the product $\eta_p\eta_m$ is specified. Furthermore, the two different optimal number of passes $k^{\text{Cl}}_{\text{opt}}$ and $k^{\text{Q}}_{\text{opt}}$ for classical or quantum MP strategies are determined by the rest of the parameters. Because of this, we can investigate the ratio of achievable precisions using optimal quantum and classical states in multi-pass schemes as a function of only $\eta$, $\eta_r$ and $\eta_p\eta_m$. This ratio is the same for both SM and TM phase estimation because the information gained differs by a factor of four for both the classical and quantum strategies.

\begin{figure}
 \includegraphics[width=0.75\columnwidth]{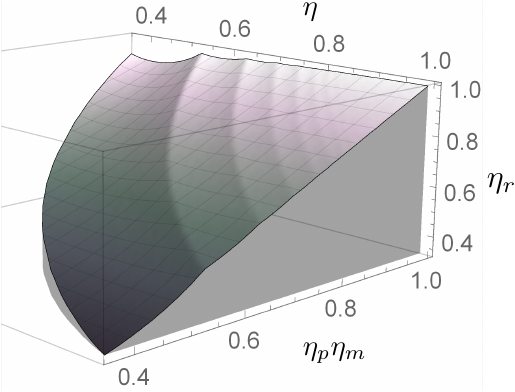}
 \caption{Plot showing which combinations of parameters $\eta,\eta_r$ and $\eta_p\eta_m$ can lead to a quantum advantage greater than that of the ideal case. The shaded gray region below the plotted surface is the region which allows a quantum advantage that reduces the RMSE by more than $20\%$ over a comparable classical strategy. Bumps in the surface correspond to the classical or quantum strategies considered with different optimal numbers of passes.}
 \label{more_loss}
\end{figure}

To establish if considering additional losses made the quantum strategy more or less advantageous relative to a classical strategy we sought to find which combinations of $\eta$, $\eta_r$ and $\eta_p\eta_m$ lead to a greater quantum advantage than in the ideal case. For a grid of $150\times150$ values of $(\eta_r,\eta_p\eta_m)$ we tried to find the highest value of $\eta_r$ such that the quantum strategy provided a reduction in RMSE of more than $20\%$. This identifies a region of the parameter space within which it may be possible to achieve a reduction in RMSE of more than $20\%$. Our method for finding the highest $\eta_r$, for each combination of $(\eta_r,\eta_p\eta_m)$, was (starting with $\eta_r=1$):
\begin{enumerate}
\item Optimize the number of passes for the classical and then the quantum strategies.
\item Then calculate the quantum advantage.
\item Iterate 1. and 2. for decreasing values of $\eta_r$ until the advantage becomes greater than a $20\%$ reduction in RMSE.
\end{enumerate}
The highest $\eta_r$ required was determined to a precision of $\pm0.00008$. A surface of these values for $\eta_r$ are plotted in Fig.~\ref{more_loss}.

From these optimizations we note that a qualitative change to the quantum strategies occurs when additional losses were included: The optimal number of passes $k^{\text{Q}}_{\text{opt}}$ is not always one as it is in the ideal case. This means that imperfect quantum strategies can also make use of multi-pass techniques.
As can be seen in this plot, the resultant surface appears to be approximately $\eta_r = \eta_p\eta_m$ where it is present. in fact, the surface lies below $\eta_r = \eta_p\eta_m$ at all points meaning that for all the 22500 values of $\eta_r$ and $\eta_p\eta_m$ we studied: in order to obtain a reduction in RMSE of more than $20\%$ the individual round-trip loss $1-\eta_r$ must be greater than the combined state preparation and measurement loss $1- \eta_p\eta_m$. in places the surface is missing which means there was no value of $\eta_r$ for which quantum techniques granted a reduction in RMSE of more than $20\%$. This will occur when it is not necessary to use more than one passage through the phase since in this case the round-trip loss will not play a role.

We have not studied the effects of imperfections in the more general optical setups of Appendix C and E. We expect that the efficacy of these setups to be more heavily degraded by component imperfection and will only offer an advantage when the sample introduces the dominant source of loss. We do report one observation about the general setups: The QFI obtained using the setups specified in Table \ref{opt_table} is almost entirely ($\geq99\%$) in the mode containing the lossy phase. Therefore, we would not expect losses in the reference mode to significantly reduce the efficacy the setups specified in Table \ref{opt_table}.

\end{document}